\documentclass[sigconf]{acmart}

\usepackage{array}
\usepackage{colortbl}
\usepackage{xcolor}

\copyrightyear{2026}
\acmYear{2026}
\setcopyright{cc}
\setcctype{by}
\acmConference[MRAC '26]{4th International Workshop on Multimodal, Generative and Responsible Affective Computing}{November 10--14, 2026}{Rio de Janeiro, Brazil}
\acmBooktitle{4th International Workshop on Multimodal, Generative and Responsible Affective Computing (MRAC '26), November 10--14, 2026, Rio de Janeiro, Brazil}
\acmDOI{10.1145/3840474.3840521}
\acmISBN{979-8-4007-2928-7/2026/11}

\begin{document}

\title{Learning to Prefer Reliably: Error-Augmented Emotion Preference Optimization with Calibrated Fusion}

\author{Zilong Huang}
\affiliation{%
  \institution{Hong Kong Polytechnic University}
  \city{Hong Kong SAR}
  \country{China}}
\email{zi-long.huang@connect.polyu.hk}

\author{Junyi Peng}
\affiliation{%
  \institution{Brno University of Technology}
  \city{Brno}
  \country{Czech Republic}}
\email{pengjy@fit.vut.cz}

\author{Junjie Li}
\affiliation{%
  \institution{Hong Kong Polytechnic University}
  \city{Hong Kong SAR}
  \country{China}}
\email{junjie98.li@connect.polyu.hk}

\author{Kai Li}
\affiliation{%
  \institution{Tsinghua University}
  \city{Beijing}
  \country{China}}
\email{li-k24@mails.tsinghua.edu.cn}

\author{Wenze Ren}
\affiliation{%
  \institution{National Taiwan University}
  \city{Taipei}
  \country{Taiwan}}
\email{d14945014@ntu.edu.tw}

\author{Kong Aik Lee}
\authornote{Corresponding author.}
\affiliation{%
  \institution{Hong Kong Polytechnic University}
  \city{Hong Kong SAR}
  \country{China}}
\email{kong-aik.lee@polyu.edu.hk}

\author{Man-Wai Mak}
\affiliation{%
  \institution{Hong Kong Polytechnic University}
  \city{Hong Kong SAR}
  \country{China}}
\email{enmwmak@polyu.edu.hk}

\author{Tatsuya Kawahara}
\affiliation{%
  \institution{Kyoto University}
  \city{Kyoto}
  \country{Japan}}
\email{tatsuya@i.kyoto-u.ac.jp}

\renewcommand{\shortauthors}{Huang et al.}

\begin{abstract}

Emotion preference learning uses pairwise comparisons between candidate descriptions to align multimodal large language models (MLLMs) with human judgments of open-ended emotion descriptions and to train reward models that capture human emotional preferences. However, conventional pairwise supervision is often sparse, typically providing only a single negative description for each positive description, and therefore offers limited coverage of the diverse ways in which an emotion description can be incorrect. In particular, models may be insufficiently exposed to semantically fluent but emotionally inconsistent descriptions. Beyond this data-level limitation, relying on a single MLLM judge introduces a distinct model-level concern: its judgments can be affected by model-specific biases when interpreting fine-grained or ambiguous multimodal emotional cues. To address these limitations, we propose \textbf{Error-Augmented Preference Optimization (EAPO)}, a framework for improving the reliability of MLLM-based emotion preference judgment at both the data and model levels. First, we construct an error-augmented dataset by generating multiple controlled and emotion-aware negative descriptions from each preferred description. We then adapt multiple independent MLLM judges to this richer supervision and aggregate their preference margins using margin-calibrated soft fusion, which maps heterogeneous margins to a common scale before aggregation. Experiments on the MER2026-EmoPrefer Challenge dataset and our error-augmented dataset demonstrate that EAPO improves emotion preference prediction and enhances the robustness of MLLM judges when evaluating fluent descriptions that conflict with the video's multimodal emotional evidence. Our code is available at \url{https://github.com/slash1028/EAPO-EmoPrefer}.
\end{abstract}



\maketitle

\section{Introduction}

Descriptive multimodal emotion recognition represents affective states in free-form language rather than with a fixed emotion label. Such descriptions can express temporal dynamics, intensity, uncertainty, and evidence distributed across facial behavior, vocal delivery, linguistic content, and scene context~\cite{lian2026emoprefer,lian2026mer}. This flexibility complicates evaluation: a single reference description cannot enumerate every defensible interpretation, and lexically similar descriptions may still conflict with the emotional evidence in the video.

Emotion preference learning provides an alternative for evaluating open-ended emotion descriptions. The MER2026 Challenge formalizes this setting in the MER-Prefer track. Given a video and two candidate descriptions, annotators compare the candidates and select the one that better reflects the observed emotional state~\cite{lian2026mer}. These comparisons support both the evaluation of descriptive multimodal emotion recognition systems and the preference-based alignment of affective models~\cite{christiano2017deep,lian2026emoprefer}.

However, this formulation has two important limitations. First, conventional pairwise preference supervision is often sparse and under-specified. Each preferred description is typically paired with only one naturally occurring rejected description. Although this rejected description identifies which candidate is less preferred, it does not explicitly indicate why it is unreliable or ensure coverage of the different error modes that may occur. Such errors may involve \textit{Emotion Flip}, \textit{Intensity Mismatch}, \textit{Evidence Contradiction}, or \textit{Modality Omission}. Consequently, preference judges may be insufficiently exposed to semantically plausible but emotionally inconsistent descriptions. Second, relying on a single MLLM judge makes the final decision dependent on that model's particular biases and failure patterns, especially when interpreting fine-grained or ambiguous multimodal emotional cues. These observations motivate a framework that improves both the coverage of error-relevant preference data and the diversity of judgment sources.

\begin{figure*}[t]
\centering
\includegraphics[width=\textwidth]{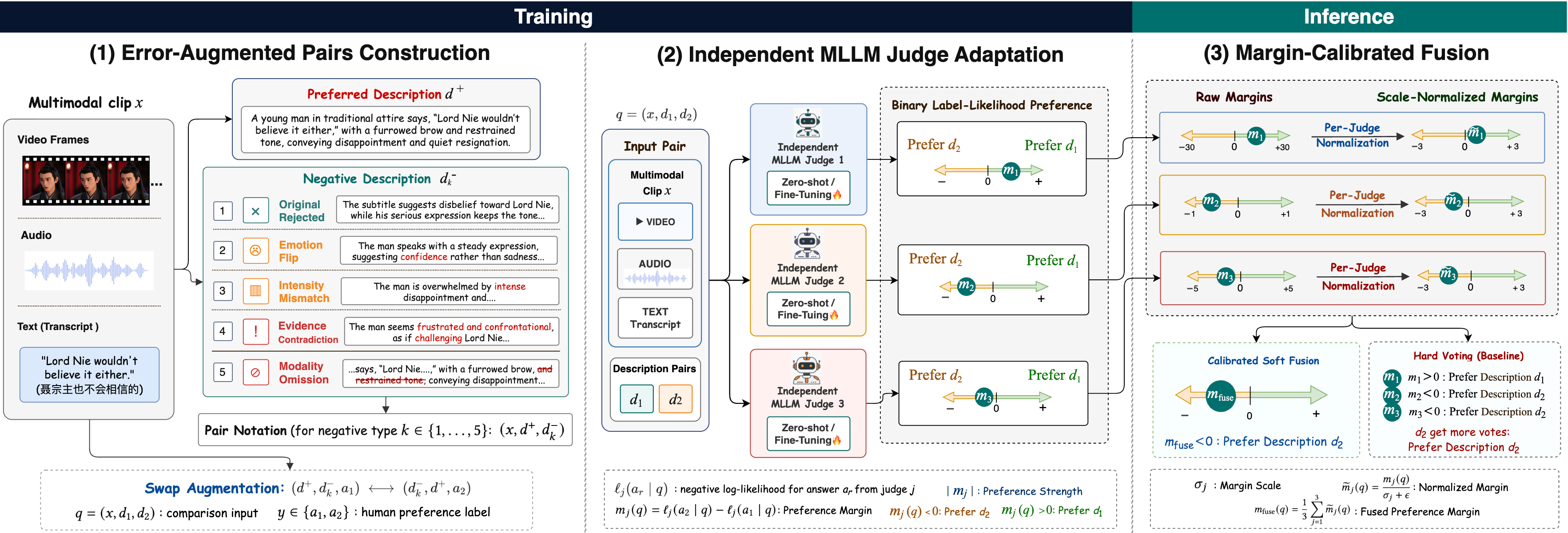}
\caption{Overview of our proposed EAPO framework. First, the error-augmented pair construction retains the
original human-rejected description and introduces 4 controlled error types. Candidate-order swapping is used during LoRA-based SFT and subsequent DPO, with each MLLM trained independently. At the inference step, signed preference margins from different MLLMs are normalized to a common scale and combined through calibrated soft fusion to obtain the final preference.}
\Description{The framework constructs five negative-pair categories from an original human preference pair, applies candidate-order swapping, and independently adapts multiple preference judges through SFT and DPO. At inference, the judges produce signed margins that are normalized to comparable scales and combined by soft fusion.}
\label{fig:framework}
\end{figure*}

In this paper, we propose \textbf{Error-Augmented Preference Optimization (EAPO)}, illustrated in Figure~\ref{fig:framework}, to improve the reliability of MLLM-based emotion preference judgment at the data and model levels. EAPO retains \textit{Original Rejected} descriptions as naturally occurring negatives and augments them with four controlled error categories: \textit{Emotion Flip}, \textit{Intensity Mismatch}, \textit{Evidence Contradiction}, and \textit{Modality Omission}. We independently adapt multiple MLLM judges to this richer supervision through supervised fine-tuning followed by standard direct preference optimization. Each judge produces an independent preference signal and does not communicate with other judges. Once multiple signals are available, we combine them through margin-calibrated soft fusion. Specifically, judge-specific signed preference margins are mapped to a common scale before aggregation, preserving the graded strength of the preference signals that would be discarded by hard voting. EAPO therefore combines naturally occurring and controlled error supervision with independent multi-judge assessment and calibrated decision-level aggregation.

Our main contributions are as follows:
\begin{itemize}
    \setlength{\itemsep}{0pt}
    \setlength{\topsep}{2pt}
    \setlength{\parsep}{0pt}

    \item We propose \textbf{Error-Augmented Preference Optimization (EAPO)}, a framework that improves MLLM-based emotion preference prediction by addressing sparse negative supervision and dependence on a single judge.

    \item We construct an error-augmented preference dataset that retains \textit{Original Rejected} descriptions and adds four controlled error types, including \textit{Modality Omission}. The resulting data are used to adapt MLLM preference judges through LoRA-based supervised fine-tuning and direct preference optimization, with candidate-order swapping to reduce position bias.

    \item We introduce margin-calibrated soft fusion for multi-judge preference prediction. The method maps preference margins from multiple independently adapted MLLM judges to a common scale before aggregation, thereby retaining graded preference evidence rather than relying on hard voting or direct averaging of raw margins.

    \item Experiments on the MER2026-EmoPrefer Challenge dataset and our controlled error-augmented subsets demonstrate the effectiveness of EAPO. Our submission ranked \textbf{sixth} on the MER2026 leaderboard and outperformed the official baseline models on the reported evaluation metrics.
\end{itemize}

\section{Related Work}

\subsection{From Emotion Recognition to Emotion Understanding}

Multimodal emotion recognition is important for affective human--computer interaction, where systems need to perceive and respond to human affect from multimodal signals~\cite{multili,xie2021robust,huang2026emoeus,yang2024prompt,wang2024enhancing}. Traditional studies mainly formulate emotion recognition as categorical or dimensional prediction, combining facial behavior, vocal delivery, linguistic content, and conversational context through robust representation learning and cross-modal fusion~\cite{xie2021robust,zhang2020emotion,yang-2023-self-scmm,yuan2023rba,huang2024mm,gu2025information,huang2026eiiscl}. Although these approaches provide effective closed-set predictions, compact labels convey limited information about the supporting evidence, emotional intensity, and ambiguity behind a decision~\cite{kang2025beyond,lian2023explainable}.

Recent studies have extended this paradigm toward open vocabulary emotion understanding, where multimodal models express affective states and their supporting cues in natural language~\cite{lian2023mer,lian2024mer,lian2025mer,lian2026mer}. Compared with categorical labels, free-form descriptions can represent subtler emotions and richer acoustic, visual, and linguistic evidence~\cite{lian2023explainable,cheng2024emotionllama}. However, free-form emotion descriptions are difficult to evaluate because the same emotional interpretation can be expressed in many valid ways. Comparing a prediction with only one reference may therefore penalize accurate descriptions that use different wording or levels of detail~\cite{lian2026emoprefer}. Emotion preference judgments address this problem by evaluating descriptions comparatively, providing a practical connection between open-ended emotion understanding and measurable output quality.

\subsection{Emotion Preference Learning}

Emotion preference learning determines which candidate emotion description is better supported by the multimodal content. This relative formulation reduces dependence on a unique textual reference and provides comparative supervision for training emotion-aware judges and reward models~\cite{kim2024emotionpreference,gao2025emodpo}. EmoPrefer establishes this setting using human-annotated preference pairs and evaluates the agreement with human preference~\cite{lian2026emoprefer}.

Existing approaches commonly employ MLLMs as preference judges. They obtain judgments through direct comparison, intermediate video descriptions, external-LLM reasoning, or model-based crowdsourcing~\cite{lian2026emoprefer,chen2024mllmjudge}. Beyond zero-shot prompting, supervised fine-tuning (SFT) can adapt an MLLM to the pairwise decision format and task-specific preference criteria, while Direct Preference Optimization (DPO) further increases the relative likelihood of preferred responses over rejected ones~\cite{rafailov2023direct}. However, conventional preference pairs typically contain only one rejected description without specifying why it is less reliable, providing limited supervision about distinct failure modes. This limitation is particularly relevant to multimodal emotion understanding, where a fluent description may misidentify the emotion or its intensity, introduce unsupported evidence, or favor one modality when acoustic, visual, and linguistic cues disagree~\cite{xing2026emotionhallucer,wang2025audio,gao2026beyond}. In addition, judgments may vary across MLLM evaluators because of model-specific biases~\cite{wang2024unfair,tan2025judgebench}. These limitations suggest that robust emotion preference learning would benefit from preference supervision covering diverse and well-defined error types, together with decision mechanisms that exploit complementary signals from multiple judges.



\section{Method}

\subsection{Problem Formulation}

Each sample consists of a comparison input
$q=(x,d_1,d_2)$ and a human preference label
$y\in\{a_1,a_2\}$, where $x$ denotes the multimodal evidence associated
with a video, including visual, acoustic, and linguistic information when
available. The two candidate emotion descriptions are denoted by $d_1$
and $d_2$. Specifically, $a_1$ indicates that $d_1$ is preferred,
whereas $a_2$ indicates that $d_2$ is preferred. Given $q$, a preference
judge predicts $\hat{y}\in\{a_1,a_2\}$. The task aims to identify the
human-preferred description by comparing both candidates against the
multimodal evidence conveyed by the video.

\subsection{Error-Augmented Negative Construction}

\begin{figure*}[!t]
    \centering
    \includegraphics[width=0.995\textwidth]{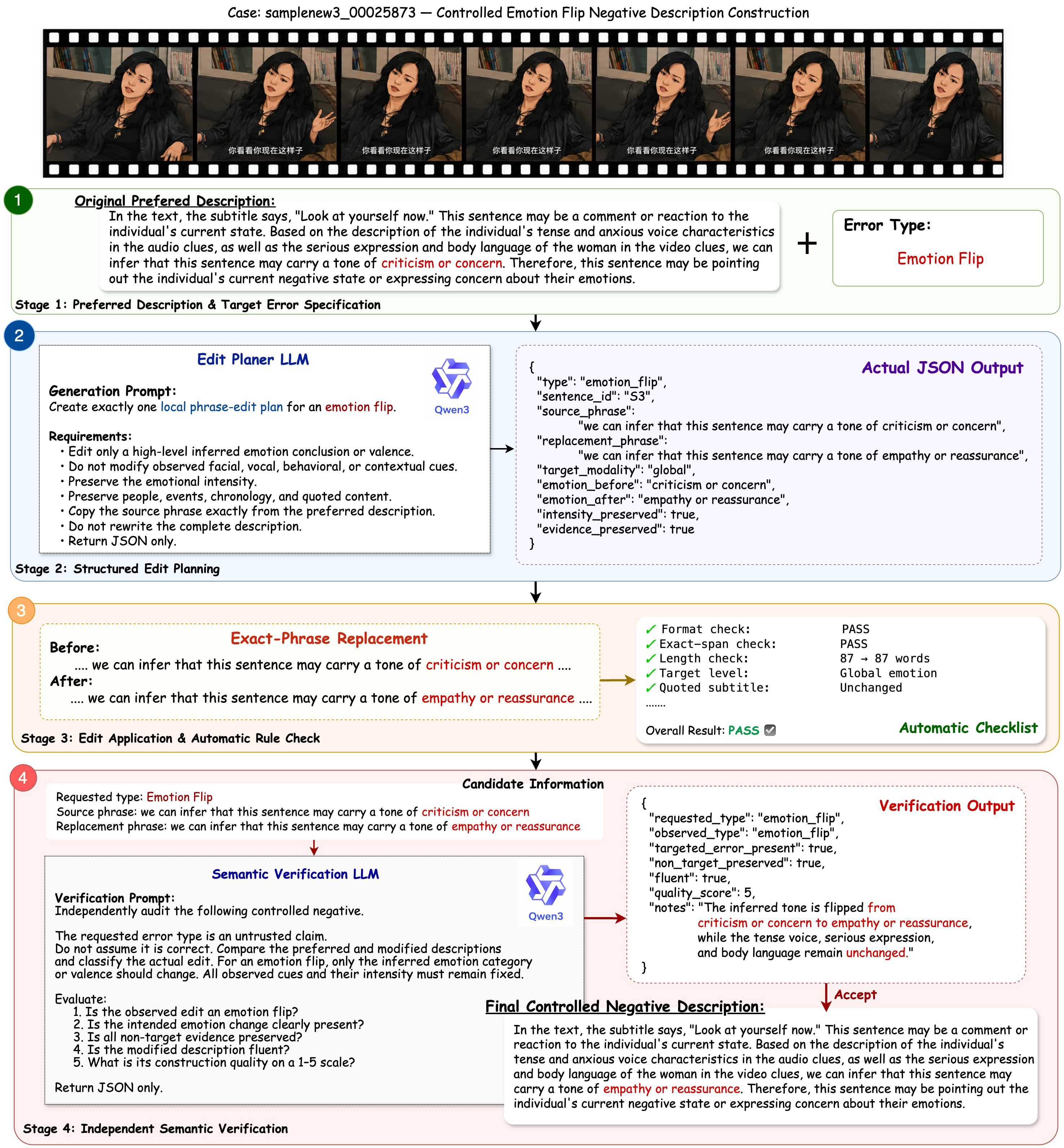}
    \caption{Illustration of the controlled Emotion Flip negative construction pipeline. Given a preferred description and a target error type, an edit-planning LLM first produces a structured local edit that alters only the high-level affective interpretation while preserving textual references to the original multimodal evidence. The proposed edit is then deterministically applied and subjected to automatic rule-based checks for structural validity and non-target preservation. Finally, an independent semantic verification LLM audits the resulting candidate for the intended error type, preservation of non-target content, fluency, and overall construction quality before it is accepted as the final controlled negative description.}
    \Description{A four-stage case study showing a preferred video description and requested Emotion Flip, a structured edit from the text-only Qwen3 Edit Planner LLM, deterministic edit application with automatic rule-based validation, and an independent semantic verification pass producing the final controlled negative.}
    \label{fig:case_study}
\end{figure*}

Preference judges are commonly trained on naturally occurring preference
pairs, which may provide limited coverage of the errors that can appear in
candidate emotion descriptions. Recent studies reveal complementary failure modes in multimodal models: emotion hallucinations may involve incorrect emotion categories, miscalibrated intensity, or descriptions unsupported by multimodal evidence~\cite{xing2026emotionhallucer}, while conflicting textual cues may override valid acoustic evidence~\cite{wang2025audio,gao2026beyond}. 
These observations motivate augmenting naturally occurring negatives with
targeted errors whose type and location can be explicitly controlled.
However, unrestricted rewriting by an LLM may simultaneously alter unrelated
facts, style, or length, introducing unintended shortcuts for preference learning. We therefore construct controlled negative descriptions
through localized edits designed to perturb a specific aspect of an otherwise
preferred description while preserving non-target content.
Figure~\ref{fig:case_study} illustrates this pipeline using an
\textit{Emotion Flip} example.

For each human-annotated preference pair, we use the preferred description
$d^+$ as the anchor for controlled negative construction and retain the
\textit{Original Rejected} description as a naturally occurring negative. We consider
four synthetic error types in addition to this category, as
summarized in Table~\ref{tab:error_taxonomy}. The preferred description is
first segmented into numbered sentences. For each synthetic error type,
the text-only Qwen3-30B-A3B-Instruct~\cite{yang2025qwen3} serves as the
Edit Planner LLM and is prompted independently
to produce a structured edit plan specifying the target sentence, source span,
and proposed modification. Each error type is processed in a separate call to
avoid mixed-error generation and to maintain explicit control over the
intended perturbation. Rather than allowing the model to rewrite the complete
description, a deterministic edit application step applies the proposed edit to the designated
span, leaving all text outside that span unchanged. For
\textit{Modality Omission}, the operation instead removes the selected
self-contained evidence span.

\begin{table}[t]
\centering
\caption{Five error types used for error-augmented dataset construction.}
\label{tab:error_taxonomy}
\small

\renewcommand{\arraystretch}{1.12}
\begin{tabular}{
    @{}
    >{\centering\arraybackslash}m{0.04\columnwidth}
    >{\centering\arraybackslash}m{0.24\columnwidth}
    >{\raggedright\arraybackslash}m{0.63\columnwidth}
    @{}
}
\toprule
\textbf{\#} & \textbf{Error Type} & \textbf{Error Definition} \\
\midrule

1 &
\textit{Original Rejected} &
The naturally occurring non-preferred description from the original
human-annotated preference pair. \\

2 &
\textit{Emotion Flip} &
Replace the inferred affect with a different or opposing emotion. \\

3 &
\textit{Intensity Mismatch} &
Preserve the inferred emotion while overstating or understating its intensity. \\

4 &
\textit{Evidence Contradiction} &
Introduce a claim that is directly contradicted by observable multimodal evidence. \\

5 &
\textit{Modality Omission} &
Remove a self-contained span describing concrete emotional evidence from one
modality while preserving the remaining content. \\

\bottomrule
\end{tabular}
\end{table}

Each edited candidate undergoes automatic rule-based validation for structural
validity, edit locality, and unintended changes or duplication. 
Candidates passing these checks are subsequently evaluated by a separate
text-only inference pass using the same text-only Qwen3 checkpoint as the semantic
verifier. It assesses whether the observed modification matches the requested
error type, preserves non-target content, and remains fluent and well formed.
Only candidates passing both automatic rule-based validation and semantic verification
are retained. Acceptance is performed independently for each error type, and
each accepted pair is stored in both candidate orders to mitigate position
bias, with error-type labels retained only as metadata.

As shown in Figure~\ref{fig:case_study}, the \textit{Emotion Flip} operation
replaces ``criticism or concern'' with ``empathy or reassurance'' while leaving
the surrounding descriptions of the multimodal evidence unchanged, producing
a controlled negative that differs primarily in its affective interpretation.

\subsection{Lightweight SFT-to-DPO Judge Adaptation}

After constructing the error-augmented preference pairs, we adapt each
MLLM judge independently through two successive stages: LoRA-based
supervised fine-tuning (SFT)~\cite{hu2022lora} followed by direct
preference optimization (DPO)~\cite{rafailov2023direct}. Both the original preference pairs and the generated pairs from the four controlled error types are used during adaptation. For each pair $(d^+,d^-_k)$, the preferred and
negative descriptions are placed in the two candidate positions, and
the corresponding $a_1$ or $a_2$ label is used as the
training target. We additionally swap the candidate order and update the
target label accordingly to reduce position-dependent predictions.

During SFT, each judge is optimized with the standard language-modeling
loss $\mathcal{L}_{\mathrm{SFT}}$, which maximizes the likelihood of the
correct answer label. This stage establishes the binary comparison
format and adapts the model to compare emotion descriptions according to
multimodal evidence. LoRA updates only a small set of trainable
parameters, allowing multiple MLLM judges to be adapted independently
with limited computational cost.

The DPO stage is initialized from the SFT checkpoint and optimized with
the standard preference loss $\mathcal{L}_{\mathrm{DPO}}$. For each
training instance, the label corresponding to $d^+$ is treated as the
preferred response, while the label corresponding to $d^-_k$ is treated
as the rejected response. DPO increases the relative likelihood of the
preferred label and further refines the preference boundary learned
during SFT. Together, the two stages enable the judges to learn the
emotion preference task while improving their discrimination between
human-preferred descriptions and descriptions containing the constructed
errors.

\subsection{Margin-Calibrated Multi-Judge Fusion}

We combine three independently adapted MLLM judges through Margin-Calibrated Multi-Judge Fusion. Given a comparison instance
$q=(x,d_1,d_2)$, all judges receive the same multimodal evidence $x$ and
candidate descriptions $(d_1,d_2)$. Here, $d_1$ and $d_2$ denote the two
candidate descriptions, while $a_1$ and $a_2$ denote the corresponding
answer labels. Specifically, $a_r$ indicates that the judge selects
description $d_r$, where $r\in\{1,2\}$.

For each instance, judge $j$ evaluates the two valid answer labels under
the same prompt. We denote their negative log-likelihoods by
$\ell_j(a_1\mid q)$ and $\ell_j(a_2\mid q)$, and define the signed
preference margin as
\begin{equation}
    m_j(q)
    =
    \ell_j(a_2\mid q)
    -
    \ell_j(a_1\mid q).
\end{equation}
A positive margin indicates a preference for $d_1$, whereas a negative
margin indicates a preference for $d_2$. The magnitude $|m_j(q)|$
reflects the strength of the judge-specific preference.

The numerical scales of raw margins may differ substantially across
judges, making direct averaging unreliable. For an evaluation split $\{q_n\}_{n=1}^{N}$, where $N$ is the total number of comparison instances in the current split, we estimate the standard
deviation of each judge's raw margins as
\begin{equation}
    \sigma_j
    =
    \sqrt{
        \frac{1}{N}
        \sum_{n=1}^{N}
        \left(
            m_j(q_n)-\mu_j
        \right)^2
    },
\end{equation}
where $\mu_j$ is the mean raw margin of judge $j$ over the same batch.
The statistics are computed independently for each judge using only its
raw margins, without preference labels.
We then apply the scale normalization:
\begin{equation}
    \widetilde{m}_j(q)
    =
    \frac{m_j(q)}{\sigma_j+\epsilon},
\end{equation}
where $\epsilon=10^{-6}$ ensures numerical stability. This transformation
places the margin magnitudes of different judges on comparable scales
while preserving their original preference directions.

Finally, the three normalized margins are combined with equal weights:
\begin{equation}
    m_{\mathrm{fuse}}(q)
    =
    \frac{1}{3}
    \sum_{j=1}^{3}
    \widetilde{m}_j(q).
\end{equation}
The final preference prediction is determined by the sign of the fused
margin:
\begin{equation}
    \hat{y}(q)
    =
    \begin{cases}
        a_1, & m_{\mathrm{fuse}}(q)>0,\\
        a_2, & m_{\mathrm{fuse}}(q)\leq0.
    \end{cases}
\end{equation}
Unlike hard voting, which retains only the discrete preference,
the proposed fusion also incorporates the relative strength of each
judge's preference after accounting for differences in margin scale.

\section{Experiments}

\begin{table}[!t]
\centering
\caption{Dataset composition and experimental role.}
\label{tab:dataset_composition}
\scriptsize
\setlength{\tabcolsep}{3pt}
\resizebox{\columnwidth}{!}{%
\begin{tabular}{llr}
\toprule
\textbf{Dataset} & \textbf{Role} & \textbf{\#Samples} \\
\midrule
EmoPrefer-Data-V2 & Normal Training & 1,618 \\
Error-Aug Train-Set & Error-aug Training & 2,908 \\
\midrule
EmoPrefer-Data & Normal Validation & 563 \\
Error-Aug Val-Set & Error-aug Validation & 944 \\
\midrule
MER-Prefer Test Stage-1& Test Evaluation & 379 \\
MER-Prefer Test Stage-2& Test Evaluation & 515 \\
\bottomrule
\end{tabular}
}
\end{table}

\begin{table*}[!t]
\centering
\caption{Comparison of zero-shot, adapted, and fused preference judges. Original Val denotes WAF on EmoPrefer-Data, the official validation set of original human-annotated preference pairs. Each Judge ID identifies a model together with its inference selection (S1 or S2) and optimization strategy. "Zero-shot" means no optimization on the model. F1 combines the post-optimization preference margins of Judges 11, 14, and 21 using margin-calibrated fusion. Gray columns report aggregate results. Best and second-best results are shown in \textbf{bold} and \underline{underline}, respectively.}
\label{tab:error_type_acc}
\tiny
\setlength{\tabcolsep}{2.5pt}
\renewcommand{\arraystretch}{0.95}
\resizebox{\textwidth}{!}{%
\begin{tabular}{cllrrrrrrr}
\toprule
\multicolumn{3}{c}{\textbf{Judge Configuration}} & \multicolumn{1}{c}{} & \multicolumn{4}{c}{\textbf{4 Generated-Error Subsets}} & \multicolumn{1}{c}{} & \multicolumn{1}{c}{} \\
\cmidrule(lr){1-3}\cmidrule(lr){5-8}
\textbf{ID} & \textbf{Model} & \textbf{Optimization Strategy} & \cellcolor[HTML]{ececec}{\textbf{Orig. Val}} & Emotion Flip & Intensity Mismatch & Evidence Contradiction & Modality Omission & \textbf{4-Error Avg} & \textbf{Swap Cons} \\
\midrule
1 & GPT-5.5 & S1 Zero-shot & \cellcolor[HTML]{ececec}{66.34} & -- & -- & -- & -- & \cellcolor[HTML]{ececec}{--} & -- \\
2 & GPT-5.5 Pro & S1 Zero-shot & \cellcolor[HTML]{ececec}{66.56} & -- & -- & -- & -- & \cellcolor[HTML]{ececec}{--} & -- \\
3 & MiMo-V2.5 & S1 Zero-shot & \cellcolor[HTML]{ececec}{68.03} & -- & -- & -- & -- & \cellcolor[HTML]{ececec}{--} & -- \\
4 & MiMo-V2.5 & S2 Zero-shot & \cellcolor[HTML]{ececec}{67.18} & -- & -- & -- & -- & \cellcolor[HTML]{ececec}{--} & -- \\
5 & Qwen3-Omni-30B-A3B-Thinking & S2 Zero-shot & \cellcolor[HTML]{ececec}{73.36} & 92.27 & 54.51 & 71.20 & 65.95 & \cellcolor[HTML]{ececec}{70.98} & 62.29 \\
6 & Qwen3-Omni-30B-A3B-Instruct & S2 Zero-shot & \cellcolor[HTML]{ececec}{73.43} & 87.06 & 52.29 & 63.49 & 77.75 & \cellcolor[HTML]{ececec}{70.15} & 54.13 \\
\midrule
7 & MiniCPM-o-2.6-8B & S1 Zero-shot & \cellcolor[HTML]{ececec}{61.53} & 83.97 & 47.27 & 66.26 & 80.34 & \cellcolor[HTML]{ececec}{69.46} & 53.50 \\
8 & MiniCPM-o-2.6-8B & S1 Normal SFT & \cellcolor[HTML]{ececec}{60.79} & 84.31 & 47.04 & 66.00 & 81.17 & \cellcolor[HTML]{ececec}{69.63} & 53.92 \\
9 & MiniCPM-o-2.6-8B & S1 Normal SFT+DPO & \cellcolor[HTML]{ececec}{60.74} & 84.31 & 45.34 & 65.69 & 81.16 & \cellcolor[HTML]{ececec}{69.12} & 53.28 \\
10 & MiniCPM-o-2.6-8B & S1 Error-Aug SFT & \cellcolor[HTML]{ececec}{71.90} & 93.22 & 70.67 & \textbf{79.07} & \textbf{92.74} & \cellcolor[HTML]{ececec}{83.92} & 75.32 \\
11 & MiniCPM-o-2.6-8B & S1 Error-Aug SFT+DPO & \cellcolor[HTML]{ececec}{73.89} & 94.95 & \textbf{79.63} & \underline{78.65} & 91.11 & \cellcolor[HTML]{ececec}{\textbf{86.09}} & \underline{76.27} \\
\midrule
12 & Qwen2.5-Omni-7B & S2 Zero-shot & \cellcolor[HTML]{ececec}{68.17} & 83.32 & 49.30 & 51.82 & 47.59 & \cellcolor[HTML]{ececec}{58.01} & 34.96 \\
13 & Qwen2.5-Omni-7B & S2 Normal SFT & \cellcolor[HTML]{ececec}{77.08} & 86.44 & 51.35 & 62.57 & 80.48 & \cellcolor[HTML]{ececec}{70.21} & 53.18 \\
14 & Qwen2.5-Omni-7B & S2 Normal SFT+DPO & \cellcolor[HTML]{ececec}{77.25} & 90.49 & 59.45 & 65.90 & 74.33 & \cellcolor[HTML]{ececec}{72.54} & 57.10 \\
15 & Qwen2.5-Omni-7B & S2 Error-Aug SFT & \cellcolor[HTML]{ececec}{77.15} & 94.32 & 67.13 & 68.58 & 78.49 & \cellcolor[HTML]{ececec}{77.13} & 64.41 \\
16 & Qwen2.5-Omni-7B & S2 Error-Aug SFT+DPO & \cellcolor[HTML]{ececec}{78.29} & 94.79 & 65.77 & 72.03 & 74.96 & \cellcolor[HTML]{ececec}{76.89} & 65.36 \\
\midrule
17 & Qwen3-Omni-30B-A3B-Instruct & S2 Zero-shot & \cellcolor[HTML]{ececec}{73.43} & 87.06 & 52.29 & 63.49 & 77.75 & \cellcolor[HTML]{ececec}{70.15} & 54.13 \\
18 & Qwen3-Omni-30B-A3B-Instruct & S2 Normal SFT & \cellcolor[HTML]{ececec}{75.65} & 86.25 & 40.66 & 67.73 & 78.48 & \cellcolor[HTML]{ececec}{68.28} & 56.46 \\
19 & Qwen3-Omni-30B-A3B-Instruct & S2 Normal SFT+DPO & \cellcolor[HTML]{ececec}{77.78} & 92.59 & 53.76 & 68.85 & 84.92 & \cellcolor[HTML]{ececec}{75.03} & 66.21 \\
20 & Qwen3-Omni-30B-A3B-Instruct & S2 Error-Aug SFT & \cellcolor[HTML]{ececec}{76.56} & 92.74 & 61.64 & 71.48 & 76.79 & \cellcolor[HTML]{ececec}{75.66} & 65.47 \\
21 & Qwen3-Omni-30B-A3B-Instruct & S2 Error-Aug SFT+DPO & \cellcolor[HTML]{ececec}{\underline{79.04}} & \underline{95.74} & 71.48 & 75.70 & \underline{92.73} & \cellcolor[HTML]{ececec}{83.91} & 75.74 \\
\midrule
F1 & Judges 11+14+21 & \textbf{Margin-Calibrated Fusion (EAPO)} & \cellcolor[HTML]{ececec}{\textbf{80.31}} & \textbf{96.21} & \underline{77.01} & 75.46 & 92.72 & \cellcolor[HTML]{ececec}{\underline{85.35}} & \textbf{76.38} \\
\bottomrule
\end{tabular}
}
\end{table*}

\begin{table}[!t]
\centering
\caption{Impact of error-augmented training and SFT-to-DPO adaptation on Qwen3-Omni-30B-A3B-Instruct.}
\label{tab:augmentation}
\scriptsize
\setlength{\tabcolsep}{3pt}
\resizebox{\columnwidth}{!}{%
\begin{tabular}{lrrr}
\toprule
\textbf{Setting} & \textbf{Orig. Val} $\uparrow$ & \textbf{4-Error Avg} $\uparrow$ & \textbf{Swap Cons} $\uparrow$ \\
\midrule
Zero-shot & 73.43 & 70.15 & 54.13 \\
Normal SFT & 75.65 & 68.28 & 56.46 \\
Normal SFT+DPO & 77.78 & 75.03 & 66.21 \\
Error-Aug SFT & 76.56 & 75.66 & 65.47 \\
\rowcolor[HTML]{ececec}
Error-Aug SFT+DPO & \textbf{79.04} & \textbf{83.91} & \textbf{75.74} \\
\bottomrule
\end{tabular}
}
\end{table}

\subsection{Datasets}

Table~\ref{tab:dataset_composition} summarizes all data splits. After tie removal and binary filtering, EmoPrefer-Data-V2 and EmoPrefer-Data contain 1,618 training and 563 validation pairs, respectively~\cite{lian2026mer}. Error-Aug Train-Set and Error-Aug Val-Set contain 2,908 and 944 generated pairs spanning the four controlled error types, including \textit{Modality Omission}. They exclude Original Rejected pairs and serve as controlled augmentation and diagnostic data. The official Stage~1 and Stage~2 test sets contain 379 and 515 samples, respectively.

Normal training uses only EmoPrefer-Data-V2, whereas error-augmented training adds Error-Aug Train-Set. \textbf{Orig. Val} and \textbf{4-Error Avg} are evaluated on EmoPrefer-Data and Error-Aug Val-Set, respectively; the official test sets are reserved for final evaluation.

\subsection{Experimental Setup}

We conducted all adaptation experiments using three open-source MLLMs as our primary preference judges: MiniCPM-o-2.6-8B~\cite{openbmb2025minicpmo}, Qwen2.5-Omni-7B~\cite{xu2025qwen25omni}, and Qwen3-Omni-30B-A3B-Instruct~\cite{xu2025qwen3omni}. For broader comparison, we additionally report zero-shot results from GPT-5.5 and GPT-5.5 Pro~\cite{openai2026gpt55,openai2026gpt55pro}, MiMo-V2.5~\cite{xiaomi2026mimov25}, and the Instruct and Thinking variants of Qwen3-Omni-30B-A3B. 
Following EmoPrefer~\cite{lian2026emoprefer}, we considered two inference strategies. \textbf{Strategy~1 (S1)} directly asks the preference judge to select the better candidate from the input video and the two descriptions. \textbf{Strategy~2 (S2)} first uses Qwen3-Omni-30B-A3B-Instruct to generate a detailed multimodal description of the video; the preference judge then uses this description as auxiliary evidence when comparing the candidates.

We used the weighted F1 (WAF), the challenge's official metric to evaluate preference prediction performance. \textbf{4-Error Avg} macro-averages WAF across the four generated-error subsets, while \textbf{Swap Cons} measures the consistency of the selected description identity under candidate-order swapping, averaged over the same subsets.

\subsection{Overall Results}

Table~\ref{tab:error_type_acc} shows that performance on Original Val and robustness to the more challenging controlled errors do not always improve together. For Qwen3-Omni, Normal SFT improves Original Val but slightly reduces 4-Error Avg, whereas Normal SFT+DPO improves all three metrics over zero-shot. Error-Aug SFT+DPO achieves the strongest performance across all three metrics. With error-augmented training, MiniCPM improves across all three metrics, whereas Qwen2.5 reaches its best Original Val. 

These complementary profiles motivate calibrated fusion. The selected judges combine an error-augmented MiniCPM checkpoint (Judge 11), a normally trained Qwen2.5 checkpoint (Judge 14), and an error-augmented Qwen3 checkpoint (Judge 21). Their fusion achieves the highest WAF of 80.31\% on Original Val, while maintaining 85.35\% in 4-Error Avg and 76.38\% in Swap Cons. The improvement therefore arises from complementary decision margins across judges rather than a single dominant checkpoint.

\subsection{Effect of Error-Augmented Training}

Table~\ref{tab:augmentation} isolates the effect of error-augmented training on Qwen3-Omni-Instruct. Training only on the official preference pairs produces mixed effects: Normal SFT improves Original Val but reduces 4-Error Avg, whereas Normal SFT+DPO improves all three metrics over zero-shot. Better fitting the original distribution therefore does not necessarily improve discrimination of controlled semantic errors.

For Qwen3-Omni, the benefit of generated pairs appears at the SFT stage and is strengthened in the full SFT-to-DPO pipeline. Error-Aug SFT improves all three metrics over Normal SFT, whereas EAPO achieves the best result in all three columns. This pattern is consistent with generated contrasts and DPO helping distinguish grounded descriptions from fluent but misleading alternatives.

\subsection{Error-Type Robustness and Fusion Analysis}

Table~\ref{tab:fusion} compares representative single judges with Hard Voting and Calibrated Fusion over Judges 11, 14, and 21, where Judge IDs follow Table~\ref{tab:error_type_acc}.
The single judges exhibit complementary strengths: Qwen3 performs best on Original Val, whereas MiniCPM is stronger on the generated errors and candidate-order consistency. Unlike Hard Voting, which discards margin magnitude, Calibrated Fusion averages scale-normalized continuous margins. It raises Original Val from 78.96\% to 80.31\%, 4-Error Avg from 84.44\% to 85.35\%, and Swap Cons from 74.79\% to 76.38\%. The fusion therefore improves preference prediction while retaining controlled-error robustness.

\begin{table}[!t]
\centering
\caption{Comparison of single-judge selection, hard voting, and calibrated fusion on Original Val and the generated-error subsets. Best results are shown in \textbf{bold}.}
\label{tab:fusion}
\scriptsize
\setlength{\tabcolsep}{3pt}
\resizebox{\columnwidth}{!}{%
\begin{tabular}{llrrr}
\toprule
\textbf{Selected Judge(s)} & \textbf{Model / Strategy} & \textbf{Orig. Val WAF} $\uparrow$ & \textbf{4-Error Avg} $\uparrow$ & \textbf{Swap Cons} $\uparrow$ \\
\midrule
Judge 11 & Best MiniCPM judge & 73.89 & \textbf{86.09} & 76.27 \\
Judge 16 & Best Qwen2.5 judge & 78.29 & 76.89 & 65.36 \\
Judge 21 & Best Qwen3-Omni judge & 79.04 & 83.91 & 75.74 \\
\midrule
Judges 11+14+21 & Hard Voting & 78.96 & 84.44 & 74.79 \\
Judges 11+14+21 & Calibrated Fusion (Ours) & \textbf{80.31} & 85.35 & \textbf{76.38} \\
\bottomrule
\end{tabular}
}
\end{table}

\subsection{Official Test Results}

Table~\ref{tab:official} reports WAF on the two official test stages, with Macro WAF calculated as their arithmetic mean. Since the official baselines report only Stage~1, Macro comparisons are restricted to our systems. Judge IDs refer to the evaluated configurations listed in Table~\ref{tab:error_type_acc} and the fusion systems combine Judges 11, 16, and 21.

All single judges outperform the strongest official baseline on Stage~1, while Stage~2 remains more challenging. Qwen3-Omni with Error-Aug SFT+DPO obtains the highest single-judge results on both stages and the best single-judge Macro WAF of 79.40\%. Zero-shot performs best within MiniCPM, whereas Error-Aug SFT+DPO performs best within Qwen2.5 and Qwen3-Omni, showing that adaptation gains vary across backbones and test stages.

Fusion produces a clearer trend. Retaining continuous preference margins improves over Hard Voting, while normalizing judge-specific margin scales further improves over raw averaging. Calibrated Fusion reaches 80.23\% on Macro WAF, compared with 79.40\% for the best single judge and 79.31\% for Raw Fusion. These results support both the use of confidence-bearing margins and their cross-judge calibration before aggregation.

\begin{table}[!t]
\centering
\caption{Official Stage~1 and Stage~2 test results in WAF (\%). Macro denotes the arithmetic mean of Stage~1 and Stage~2 WAF. Best results are shown in \textbf{bold}.}
\label{tab:official}
\tiny
\setlength{\tabcolsep}{2.5pt}
\resizebox{\columnwidth}{!}{%
\begin{tabular}{lllrrr}
\toprule
\textbf{System} & \textbf{Model / Judge(s)} & \textbf{Configuration} & \textbf{Stage 1} $\uparrow$ & \textbf{Stage 2} $\uparrow$ & \textbf{Macro} $\uparrow$ \\
\midrule
Official baseline & Qwen2-Audio & Zero-shot & 36.08 & -- & -- \\
Official baseline & Video-LLaVA & Zero-shot & 36.76 & -- & -- \\
Official baseline & LLaMA-VID & Zero-shot & 36.76 & -- & -- \\
Official baseline & LLaVA-Next-Video & Zero-shot & 41.31 & -- & -- \\
Official baseline & Qwen2.5-VL & Zero-shot & 76.77 & -- & -- \\
Official baseline & Qwen2.5-Omni & Zero-shot & 78.74 & -- & -- \\
\midrule
Single judge & MiniCPM-o-2.6-8B & S1 Zero-shot & 87.34 & 66.13 & 76.73 \\
Single judge & MiniCPM-o-2.6-8B & S1 Normal SFT+DPO & 86.81 & 65.92 & 76.36 \\
Single judge & MiniCPM-o-2.6-8B & S1 Error-Aug SFT+DPO & 85.97 & 65.88 & 75.92 \\
Single judge & Qwen2.5-Omni-7B & S2 Zero-shot & 83.44 & 66.97 & 75.20 \\
Single judge & Qwen2.5-Omni-7B & S2 Normal SFT+DPO & 86.55 & 66.43 & 76.49 \\
Single judge & Qwen2.5-Omni-7B & S2 Error-Aug SFT+DPO & 87.18 & 66.37 & 76.77 \\
Single judge & Qwen3-Omni-30B-A3B & S2 Zero-shot & 88.57 & 66.94 & 77.76 \\
Single judge & Qwen3-Omni-30B-A3B & S2 Normal SFT+DPO & 88.93 & 67.43 & 78.18 \\
Single judge & Qwen3-Omni-30B-A3B & S2 Error-Aug SFT+DPO & 90.25 & 68.56 & 79.40 \\
\midrule
Hard Voting & Judges 11+16+21 & Majority votes & 88.93 & 68.02 & 78.47 \\
Raw Fusion & Judges 11+16+21 & Raw-margin mean & 89.98 & 68.63 & 79.31 \\
\rowcolor[HTML]{ececec}
EAPO (Ours) & Judges 11+16+21 & Normalized mean & \textbf{91.30} & \textbf{69.17} & \textbf{80.23} \\

\bottomrule
\end{tabular}
}
\end{table}

\subsection{Limitations and Future Work}

Due to time constraints, the construction of error-aware pairs and the intermediate descriptions required by the strategy S2 relied primarily on MLLM generation, whose quality may vary across samples. Although the strategy S2 can assist preference judgment by providing richer multimodal evidence before candidate comparison, its effectiveness depends on the quality of the generated description; missing or inaccurate cues may propagate to the final decision. Fusion calibration has also been evaluated only with the current judge families and data distribution, and its stability under model replacement or domain shift remains unclear.

Future work will focus on combining cross-modal consistency verification with targeted human review to filter ambiguous pairs and improve error-type precision. We will also explore a multi-agent description-generation pipeline to produce better-grounded descriptions for the strategy S2.

\section{Conclusions}

We introduced Error-Augmented Preference Optimization \textbf{(EAPO)} for robust multimodal emotion preference learning. EAPO augments human preference pairs with four controlled error types and incorporates the augmented dataset into SFT-to-DPO adaptation. Experiments on the original validation set, the four controlled-error subsets, and the official two-stage test set show that adaptation using only the original preference pairs may improve in-distribution agreement while weakening robustness to plausible semantic errors. In contrast, training on our generated error-augmented dataset enables EAPO to improve agreement with human preferences while maintaining strong error-type robustness and candidate-order consistency. We further aggregate independently trained MLLM judges through calibrated continuous preference margins. Together, error-augmented preference training and calibrated multi-judge fusion preserve complementary information about preference strength and achieve the strongest overall performance among the evaluated configurations.

\begin{acks}
This work was supported in part by the Research Grants Council of the Hong Kong SAR (Grant No 15228223), and The Hong Kong Polytechnic University (Project ID P0049192).
\end{acks}

\bibliographystyle{ACM-Reference-Format}
\bibliography{refs}

@inproceedings{hu2022lora,
  title={{LoRA}: Low-Rank Adaptation of Large Language Models},
  author={Hu, Edward J. and Shen, Yelong and Wallis, Phillip and Allen-Zhu, Zeyuan and Li, Yuanzhi and Wang, Shean and Wang, Lu and Chen, Weizhu},
  booktitle={International Conference on Learning Representations},
  year={2022},
  url={https://openreview.net/forum?id=nZeVKeeFYf9}
}

@inproceedings{rafailov2023direct,
  title={Direct Preference Optimization: Your Language Model Is Secretly a Reward Model},
  author={Rafailov, Rafael and Sharma, Archit and Mitchell, Eric and Manning, Christopher D. and Ermon, Stefano and Finn, Chelsea},
  booktitle={Advances in Neural Information Processing Systems},
  volume={36},
  pages={53728--53741},
  year={2023},
  publisher={Curran Associates, Inc.},
  address={Red Hook, NY, USA},
  doi={10.52202/075280-2338},
  url={https://papers.nips.cc/paper_files/paper/2023/hash/a85b405ed65c6477a4fe8302b5e06ce7-Abstract-Conference.html}
}

@misc{lian2026mer,
  title={{MER 2026}: From Discriminative Emotion Recognition to Generative Emotion Understanding},
  author={Lian, Zheng and Peng, Xiaojiang and Xu, Kele and Jia, Ziyu and Che, Xinyi and Cheng, Zebang and Ma, Fei and Cui, Laizhong and Zhang, Yazhou and Liu, Xin and others},
  year={2026},
  eprint={2604.19417},
  archivePrefix={arXiv},
  primaryClass={cs.MM},
  doi={10.48550/arXiv.2604.19417},
  url={https://arxiv.org/abs/2604.19417}
}

@inproceedings{lian2026emoprefer,
  title={{EmoPrefer}: Can Large Language Models Understand Human Emotion Preferences?},
  author={Lian, Zheng and Sun, Licai and Chen, Lan and Chen, Haoyu and Cheng, Zebang and Zhang, Fan and Jia, Ziyu and Ma, Ziyang and Ma, Fei and Peng, Xiaojiang and others},
  booktitle={The Fourteenth International Conference on Learning Representations},
  year={2026},
  url={https://openreview.net/forum?id=EhA4znYsuG}
}

@misc{xu2025qwen25omni,
      title={{Qwen2.5-Omni} Technical Report}, 
      author={Jin Xu and Zhifang Guo and Jinzheng He and Hangrui Hu and Ting He and Shuai Bai and Keqin Chen and Jialin Wang and Yang Fan and Kai Dang and Bin Zhang and Xiong Wang and Yunfei Chu and Junyang Lin},
      year={2025},
      eprint={2503.20215},
      archivePrefix={arXiv},
      primaryClass={cs.CL},
      url={https://arxiv.org/abs/2503.20215}, 
}

@misc{xu2025qwen3omni,
  title={{Qwen3-Omni} Technical Report},
  author={Xu, Jin and Guo, Zhifang and Hu, Hangrui and Chu, Yunfei and Wang, Xiong and He, Jinzheng and Wang, Yuxuan and Shi, Xian and He, Ting and Zhu, Xinfa and others},
  year={2025},
  eprint={2509.17765},
  archivePrefix={arXiv},
  primaryClass={cs.CL},
  doi={10.48550/arXiv.2509.17765},
  url={https://arxiv.org/abs/2509.17765}
}

@misc{yang2025qwen3,
  title={{Qwen3} Technical Report},
  author={Yang, An and Li, Anfeng and Yang, Baosong and Zhang, Beichen and Hui, Binyuan and Zheng, Bo and others},
  year={2025},
  eprint={2505.09388},
  archivePrefix={arXiv},
  primaryClass={cs.CL},
  doi={10.48550/arXiv.2505.09388},
  url={https://arxiv.org/abs/2505.09388}
}

@inproceedings{christiano2017deep,
  title={Deep Reinforcement Learning from Human Preferences},
  author={Christiano, Paul F. and Leike, Jan and Brown, Tom B. and Martic, Miljan and Legg, Shane and Amodei, Dario},
  booktitle={Advances in Neural Information Processing Systems},
  volume={30},
  pages={4299--4307},
  year={2017},
  publisher={Curran Associates, Inc.},
  address={Red Hook, NY, USA}
}

@misc{openbmb2025minicpmo,
  author={{OpenBMB}},
  title={{MiniCPM-o 2.6}},
  year={2025},
  howpublished={Hugging Face Model Card},
  url={https://huggingface.co/openbmb/MiniCPM-o-2_6},
  note={Accessed 10 August 2026}
}

@misc{lian2023explainable,
  title={Explainable Multimodal Emotion Recognition},
  author={Lian, Zheng and Sun, Haiyang and Sun, Licai and Gu, Hao and Wen, Zhuofan and Zhang, Siyuan and Chen, Shun and Xu, Mingyu and Xu, Ke and Chen, Kang and others},
  year={2023},
  eprint={2306.15401},
  archivePrefix={arXiv},
  primaryClass={cs.MM},
  doi={10.48550/arXiv.2306.15401},
  url={https://arxiv.org/abs/2306.15401}
}

@article{cheng2024emotionllama,
  title={{Emotion-LLaMA}: Multimodal emotion recognition and reasoning with instruction tuning},
  author={Cheng, Zebang and Cheng, Zhi-Qi and He, Jun-Yan and Sun, Jingdong and Wang, Kai and Lin, Yuxiang and Lian, Zheng and Peng, Xiaojiang and Hauptmann, Alexander G},
  journal={Advances in Neural Information Processing Systems},
  volume={37},
  pages={110805--110853},
  year={2024},
  doi={10.52202/079017-3518},
  url={https://proceedings.neurips.cc/paper_files/paper/2024/hash/c7f43ada17acc234f568dc66da527418-Abstract-Conference.html}
}

@inproceedings{wang2024unfair,
  title={Large Language Models Are Not Fair Evaluators},
  author={Wang, Peiyi and Li, Lei and Chen, Liang and Cai, Zefan and Zhu, Dawei and Lin, Binghuai and Cao, Yunbo and Kong, Lingpeng and Liu, Qi and Liu, Tianyu and Sui, Zhifang},
  booktitle={Proceedings of the 62nd Annual Meeting of the Association for Computational Linguistics (Volume 1: Long Papers)},
  pages={9440--9450},
  year={2024},
  address={Bangkok, Thailand},
  publisher={Association for Computational Linguistics},
  doi={10.18653/v1/2024.acl-long.511},
  url={https://aclanthology.org/2024.acl-long.511/}
}

@inproceedings{tan2025judgebench,
  title={{JudgeBench}: A Benchmark for Evaluating {LLM}-Based Judges},
  author={Tan, Sijun and Zhuang, Siyuan and Montgomery, Kyle and Tang, William Y. and Cuadron, Alejandro and Wang, Chenguang and Popa, Raluca Ada and Stoica, Ion},
  booktitle={The Thirteenth International Conference on Learning Representations},
  year={2025},
  url={https://openreview.net/forum?id=G0dksFayVq}
}

@inproceedings{chen2024mllmjudge,
  title={{MLLM}-as-a-Judge: Assessing Multimodal {LLM}-as-a-Judge with Vision-Language Benchmark},
  author={Chen, Dongping and Chen, Ruoxi and Zhang, Shilin and Wang, Yaochen and Liu, Yinuo and Zhou, Huichi and Zhang, Qihui and Wan, Yao and Zhou, Pan and Sun, Lichao},
  booktitle={Proceedings of the 41st International Conference on Machine Learning},
  pages={6562--6595},
  year={2024},
  volume={235},
  series={Proceedings of Machine Learning Research},
  publisher={PMLR},
  address={Vienna, Austria},
  url={https://proceedings.mlr.press/v235/chen24h.html}
}

@article{xie2021robust,
  title={Robust Multimodal Emotion Recognition from Conversation with Transformer-Based Crossmodality Fusion},
  author={Xie, Baijun and Sidulova, Mariia and Park, Chung Hyuk},
  journal={Sensors},
  volume={21},
  number={14},
  pages={4913},
  year={2021},
  publisher={MDPI},
  doi={10.3390/s21144913},
  url={https://doi.org/10.3390/s21144913}
}

@article{zhang2020emotion,
  title={Emotion Recognition Using Multi-Modal Data and Machine Learning Techniques: A Tutorial and Review},
  author={Zhang, Jianhua and Yin, Zhong and Chen, Peng and Nichele, Stefano},
  journal={Information Fusion},
  volume={59},
  pages={103--126},
  year={2020},
  publisher={Elsevier},
  doi={10.1016/j.inffus.2020.01.011},
  url={https://doi.org/10.1016/j.inffus.2020.01.011}
}

@article{yuan2023rba,
  title={{RBA-GCN}: Relational Bilevel Aggregation Graph Convolutional Network for Emotion Recognition},
  author={Yuan, Lin and Huang, Guoheng and Li, Fenghuan and Yuan, Xiaochen and Pun, Chi-Man and Zhong, Guo},
  journal={IEEE/ACM Transactions on Audio, Speech, and Language Processing},
  year={2023},
  volume={31},
  pages={2325--2337},
  publisher={IEEE},
  doi={10.1109/TASLP.2023.3284509},
  url={https://doi.org/10.1109/TASLP.2023.3284509}
}

@inproceedings{yang-2023-self-scmm,
    title = "Self-adaptive Context and Modal-interaction Modeling For Multimodal Emotion Recognition",
    author = "Yang, Haozhe  and
      Gao, Xianqiang  and
      Wu, Jianlong  and
      Gan, Tian  and
      Ding, Ning  and
      Jiang, Feijun  and
      Nie, Liqiang",
    editor = "Rogers, Anna  and
      Boyd-Graber, Jordan  and
      Okazaki, Naoaki",
    booktitle = "Findings of the Association for Computational Linguistics: ACL 2023",
    month = jul,
    year = "2023",
    address = "Toronto, Canada",
    publisher = "Association for Computational Linguistics",
    url = "https://aclanthology.org/2023.findings-acl.390/",
    doi = "10.18653/v1/2023.findings-acl.390",
    pages = "6267--6281",
}

@article{kang2025beyond,
  title={Beyond Single Emotion: Multi-Label Approach to Conversational Emotion Recognition},
  author={Kang, Yujin and Cho, Yoon-Sik},
  journal={Proceedings of the AAAI Conference on Artificial Intelligence},
  volume={39},
  number={23},
  pages={24321--24329},
  year={2025},
  doi={10.1609/aaai.v39i23.34609},
  url={https://ojs.aaai.org/index.php/AAAI/article/view/34609}
}

@INPROCEEDINGS{multili,
  author={Wang, Xuechen and Zhao, Shiwan and Sun, Haoqin and Wang, Hui and Zhou, Jiaming and Qin, Yong},
  booktitle={ICASSP 2025 - 2025 IEEE International Conference on Acoustics, Speech and Signal Processing (ICASSP)}, 
  title={Enhancing Multimodal Emotion Recognition through Multi-Granularity Cross-Modal Alignment}, 
  year={2025},
  pages={1--5},
  publisher={IEEE},
  address={Hyderabad, India},
  doi={10.1109/ICASSP49660.2025.10889156},
  url={https://doi.org/10.1109/ICASSP49660.2025.10889156}}

@misc{openai2026gpt55,
  author = {{OpenAI}},
  title  = {{GPT-5.5 Model}},
  year   = {2026},
  howpublished = {OpenAI API Documentation},
  url    = {https://developers.openai.com/api/docs/models/gpt-5.5},
  note   = {Accessed 9 August 2026}
}

@misc{openai2026gpt55pro,
  author = {{OpenAI}},
  title  = {{GPT-5.5 Pro Model}},
  year   = {2026},
  howpublished = {OpenAI API Documentation},
  url    = {https://developers.openai.com/api/docs/models/gpt-5.5-pro},
  note   = {Accessed 9 August 2026}
}

@misc{xiaomi2026mimov25,
  author = {{Xiaomi MiMo Team}},
  title  = {{MiMo-V2.5}},
  year   = {2026},
  howpublished = {Hugging Face Model Card},
  url    = {https://huggingface.co/XiaomiMiMo/MiMo-V2.5}
}

@inproceedings{xing2026emotionhallucer,
  author    = {Xing, Bohao and Liu, Xin and Zhao, Guoying and Liu, Chengyu
               and Fu, Xiaolan and K{\"a}lvi{\"a}inen, Heikki},
  title     = {{EmotionHallucer}: Evaluating Emotion Hallucinations in
               Multimodal Large Language Models},
  booktitle = {The Fourteenth International Conference on Learning Representations},
  year      = {2026},
  url       = {https://openreview.net/forum?id=ahWmeQG3K2}
}

@inproceedings{wang2025audio,
  author    = {Wang, Cheng and Deng, Gelei and Yang, Xianglin and Qiu, Han
               and Zhang, Tianwei},
  title     = {When Audio and Text Disagree: Revealing Text Bias in
               Large Audio-Language Models},
  booktitle = {Proceedings of the 2025 Conference on Empirical Methods
               in Natural Language Processing},
  pages     = {4878--4888},
  year      = {2025},
  month     = nov,
  address   = {Suzhou, China},
  publisher = {Association for Computational Linguistics},
  doi       = {10.18653/v1/2025.emnlp-main.246},
  url       = {https://aclanthology.org/2025.emnlp-main.246/}
}

@misc{gao2026beyond,
  author = {Gao, Yichen and Zhang, Yiqun and Wang, Zijing and Li, Yujia
            and Guo, Heng and Wu, Xi and Yang, Xiaocui and Feng, Shi
            and Zhang, Yifei and Wang, Daling},
  title  = {Beyond Text Following: Repairable Arbitration Reversals in
            Audio-Language Models},
  year   = {2026},
  eprint = {2606.05161},
  archivePrefix = {arXiv},
  url    = {https://arxiv.org/abs/2606.05161}
}

@inproceedings{lian2023mer,
  title={{MER 2023}: Multi-Label Learning, Modality Robustness, and Semi-Supervised Learning},
  author={Lian, Zheng and Sun, Haiyang and Sun, Licai and Chen, Kang and Xu, Mingyu and Wang, Kexin and Xu, Ke and He, Yu and Li, Ying and Zhao, Jinming and others},
  booktitle={Proceedings of the 31st ACM International Conference on Multimedia},
  pages={9610--9614},
  year={2023},
  publisher={Association for Computing Machinery},
  address={New York, NY, USA},
  doi={10.1145/3581783.3612836},
  url={https://doi.org/10.1145/3581783.3612836}
}

@inproceedings{lian2024mer,
  title={{MER 2024}: Semi-Supervised Learning, Noise Robustness, and Open-Vocabulary Multimodal Emotion Recognition},
  author={Lian, Zheng and Sun, Haiyang and Sun, Licai and Wen, Zhuofan and Zhang, Siyuan and Chen, Shun and Gu, Hao and Zhao, Jinming and Ma, Ziyang and Chen, Xie and others},
  booktitle={Proceedings of the 2nd International Workshop on Multimodal and Responsible Affective Computing},
  pages={41--48},
  year={2024},
  publisher={Association for Computing Machinery},
  address={New York, NY, USA},
  doi={10.1145/3689092.3689959},
  url={https://doi.org/10.1145/3689092.3689959}
}

@inproceedings{lian2025mer,
  title={{MER 2025}: When Affective Computing Meets Large Language Models},
  author={Lian, Zheng and Liu, Rui and Xu, Kele and Liu, Bin and Liu, Xuefei and Zhang, Yazhou and Liu, Xin and Li, Yong and Cheng, Zebang and Zuo, Haolin and others},
  booktitle={Proceedings of the 33rd ACM International Conference on Multimedia},
  pages={13837--13842},
  year={2025},
  publisher={Association for Computing Machinery},
  address={New York, NY, USA},
  doi={10.1145/3746027.3762007},
  url={https://doi.org/10.1145/3746027.3762007}
}

@misc{huang2026eiiscl,
      title={EII-SCL: Harnessing Emotional Inertia for Multimodal Emotion Recognition in Conversation}, 
      author={Zilong Huang and Kong Aik Lee and Chong-Xin Gan and Zezhong Jin and Ruichen Zuo and Man-Wai Mak},
      year={2026},
      eprint={2607.17366},
      archivePrefix={arXiv},
      primaryClass={cs.MM},
      url={https://arxiv.org/abs/2607.17366}, 
}

@misc{huang2026emoeus,
      title={EmoEUS: Uncertainty Supervision for Multimodal Emotion Recognition in Conversation}, 
      author={Zilong Huang and Kong Aik Lee and Junjie Li and Zhe Li and Man-Wai Mak},
      year={2026},
      eprint={2607.18336},
      archivePrefix={arXiv},
      primaryClass={cs.MM},
      url={https://arxiv.org/abs/2607.18336}, 
}

@inproceedings{huang2024mm,
  title={{MM-NodeFormer}: Node Transformer Multimodal Fusion for Emotion Recognition in Conversation},
  author={Huang, Zilong and Mak, Man-Wai and Lee, Kong Aik},
  booktitle={Interspeech 2024},
  pages={4069--4073},
  year={2024},
  publisher={ISCA},
  address={Kos Island, Greece},
  doi={10.21437/Interspeech.2024-538},
  url={https://www.isca-archive.org/interspeech_2024/huang24i_interspeech.html}
}

@misc{kim2024emotionpreference,
  title   = {Empathetic Response in Audio-Visual Conversations Using
             Emotion Preference Optimization and MambaCompressor},
  author  = {Kim, Yeonju and Park, Se Jin and Ro, Yong Man},
  year    = {2024},
  eprint  = {2412.17572},
  archivePrefix = {arXiv},
  primaryClass = {cs.MM},
  doi     = {10.48550/arXiv.2412.17572},
  url     = {https://arxiv.org/abs/2412.17572}
}

@inproceedings{gao2025emodpo,
  title     = {{Emo-DPO}: Controllable Emotional Speech Synthesis through
               Direct Preference Optimization},
  author    = {Gao, Xiaoxue and Zhang, Chen and Chen, Yiming and
               Zhang, Huayun and Chen, Nancy F.},
  booktitle = {2025 IEEE International Conference on Acoustics,
               Speech and Signal Processing (ICASSP)},
  year      = {2025},
  pages     = {1--5},
  publisher = {IEEE},
  address   = {Hyderabad, India},
  doi       = {10.1109/ICASSP49660.2025.10888737},
  url       = {https://arxiv.org/abs/2409.10157}
}

@article{gu2025information,
  author = {Tiquan Gu and Zhenzhen He and Zhe Li and Yaling Wan},
  title = {Information-Assisted and Sentiment Relation-Driven for Aspect-Based Sentiment Analysis},
  journal = {Expert Systems with Applications},
  year = {2025},
  volume = {278},
  pages = {127308},
  publisher = {Elsevier},
}

@inproceedings{yang2024prompt,
  author = {Dan Yang and Xiuhong Li and Zhe Li and Chenyu Zhou and Xiaofan Wang and Fan Chen},
  title = {Prompt Fusion Interaction Transformer for Aspect-Based Multimodal Sentiment Analysis},
  booktitle = {Proceedings of the 2024 IEEE International Conference on Multimedia and Expo (ICME)},
  year = {2024},
  pages = {1--6},
  address = {Niagara Falls, ON, Canada},
  publisher = {IEEE}
}

@inproceedings{wang2024enhancing,
  author = {Xiaofan Wang and Xiuhong Li and Zhe Li and Chenyu Zhou and Fan Chen and Dan Yang},
  title = {Enhancing Cross-Modal Alignment in Multimodal Sentiment Analysis via Prompt Learning},
  booktitle = {Proceedings of the 9th Chinese Conference on Pattern Recognition and Computer Vision (PRCV)},
  year = {2024},
  pages = {541--554},
  address = {Urumqi, China},
  publisher = {Springer},
  series = {Lecture Notes in Computer Science},
  volume = {15413}
}

\end{document}